\documentclass{article}
\usepackage{authblk}
\usepackage{graphicx}
\usepackage{natbib}
\usepackage{color,colortbl,xcolor}
\usepackage{amsmath}
\usepackage{amssymb}
\usepackage{latexsym}
\usepackage{morefloats}
\usepackage{multirow}
\usepackage{soul}

\definecolor{LightCyan}{rgb}{0.88,1,1}

\newsavebox{\astrutbox}
\sbox{\astrutbox}{\rule[-5pt]{0pt}{20pt}}

\date{}
\title{Comparison of turbulence profiles in high Reynolds number turbulent boundary layers and validation of a predictive model.}

\author[1]{J.-P. LAVAL}
\author[2]{J.C. VASSILICOS}
\author[3]{J.-M. FOUCAUT}
\author[3]{M. STANISLAS}

\affil[1]{CNRS, FRE 3723 -LML- Laboratoire de Mécanique de Lille, F-59000 Lille, France}
\affil[2]{Department of Aeronautics, Imperial College London, London SW7 2AZ, United Kingdom}
\affil[3]{Centrale Lille, FRE 3723 -LML- Laboratoire de Mécanique de Lille, F-59000 Lille, France}
\affil[4]{Centrale Lille, FRE 3723 -LML- Laboratoire de Mécanique de Lille, F-59000 Lille, France}

\begin{document}

\maketitle

\begin{abstract}
The modified Townsend-Perry attached eddy model of
Vassilicos et al (2015) combines the outer peak/plateau behaviour of
rms streamwise turbulence velocity profiles and the Townsend-Perry
log-decay of these profiles at higher distances from the wall. This
model was validated by these authors for high Reynolds number
turbulent pipe flow data and is shown here to describe equally well and
with about the same parameter values turbulent boundary layer flow
data from four different facilities and a wide range of Reynolds
numbers. The model has predictive value as, when extrapolated to the
extremely high Reynolds numbers of the SLTEST data obtained at the
Great Salt Lake Desert atmospheric test facility, it matches these
data quite well.\\
\end{abstract}

 \section{Introduction}

The structure of zero pressure gradient turbulence
boundary layer (TBL) flows has been a
subject of both fundamental and applied research for many decades. 
Townsend (1976), Perry \& Abel (1979) and Perry et
al (1986) developed the well-known attached-eddy model
 to predict the profile of the turbulent
kinetic energy with distance from the wall. This model is operative  in an intermediate 
range far from the viscous layer and predicts that the turbulent kinetic energy scales
 with the square of the wall friction velocity $u_\tau$ and decreases logarithmically 
with distance to the wall. The model leads to a logarithmic decay of the mean square 
streamwise and spanwise fluctuating velocity as function of the wall distance $y$:
\begin{equation}
  \frac{\overline{{u^\prime}^2}}{u_\tau^2} = B_1 - A_1 log\left( \frac{y}{\delta}\right)
  \label{eq:logdecay}
\end{equation}
where $A_1$ is the Townsend-Perry
constant, $B_1$  is the additive constant for the variance and $\delta$ is
the pipe radius, channel half width of turbulent boundary layer
thickness. The logarithmic decay of the mean square streamwise velocity has been 
characterised for several high Reynolds turbulent boundary layer flows
 (see e.g. \cite{marusic13,vincenti13}).
\cite{vallikivi15} compared the behavior of $\overline{u'^{2}}$ in turbulent
boundary layers and in turbulent pipe flows and concluded  as \cite{marusic13}
 that the constant of the logarithm decay of Eq. (\ref{eq:logdecay}) are very 
similar in both flows ($A_1=1.24$, $B_1=1.48$).

Various measurements of  turbulent boundary layers and turbulent pipe flows 
over the last 20 years show that an outer peak or plateau appears in the mean 
square fluctuating streamwise velocity above the buffer region  at distances 
which depend on the Reynolds number. This outer peak is known to be the consequence
of very large scale motions which develop at high Reynolds numbers 
(see \cite{hultmark12,hultmark13}, \cite{vallikivi15}, \cite{vincenti13}).
Despite several experimental studies of turbulent boundary layers flows at 
fairly large Reynolds numbers ($Re_\tau>7\,000$), there is consensus
neither on the slope of the log decay of streamwise normal Reynolds stresses
nor on the increase of an outer peak and its wall normal distance for very 
large Reynolds number. 
This outer peak or plateau cannot be explained by
the Townsend-Perry attached eddy model. This model leads to no more than
(\ref{eq:logdecay})  throughout the mean flow's logarithm range.
 \cite{vassilicos15} proposed a new model 
for the streamwise velocity spectra including a $k^{-m}$ slope (with $0<m<1$)
 in a range of scales larger than those responsible for the $k^{-1}$ slope
 of the attached eddy model of \cite{perry86}.
This model, which can fit the Princeton pipe flow
data over a range of Reynolds numbers up to the maximum available
$Re_{\tau}= 98\,000$, does not only predict the amplitude and position of
the outer peak of the mean square fluctuating streamwise velocity
profiles but also supports a more realistic variation of the integral
length scale with distance to the wall than that predicted by the
Townsend-Perry model.
If most studies agree on the logarithmic decay of  ${\overline{{u^\prime}^2}}$
the scaling of the outer peak is less consensual. The advantage of the 
model proposed by \cite{vassilicos15} is to  combine both the outer peak
behavior and the logarithm decaying range in a single model. 
In this paper we apply this model of Vassilicos et al (2015) to high 
Reynolds number turbulent boundary layer data from various independent 
facilities around the world and establish that the model fits such data too
without much variation in fitting parameters.

\section{The modified Townsend-Perry model}

The modified model proposed by \cite{vassilicos15} for the  energy spectrum 
$E_{11}(k_{1},y)$ in the region $\nu/u_{\tau} \ll y \ll \delta$ is defined in
four wavenumber ranges. The Kolomogorov range $1/y <k_{1}$ is identical to 
the Townsend-Perry model following the form $E_{11} (k_{1}, y) \sim \epsilon^{2/3}
 k_{1}^{-5/3} g_{K} (k_{1}y, k_{1}\eta)$. However, what is referred to as the
``attached eddy'' range where $E_{11}(k_{1}) \approx C_{0} u_{\tau}^{2} k_{1}^{-1}$ in
the Towsend-Perry model is restricted to the range $1/\delta_{*} < k_{1} <1/y$ 
(instead of the  Towsend-Perry model's original range $1/\delta < k_{1} <1/y$)
and a new scaling $E_{11}(k_{1}) \approx C_{1} u_{\tau}^{2} \delta (k_{1}\delta)^{-m}$
 with $0<m<1$ is proposed in the range $1/\delta_{\infty} < k_{1} <1/\delta_{*}$ 
where $\delta_{*}$ and $\delta_{\infty}$ are two large length-scales such that 
$\delta_{*} \le \delta_{\infty}$. The spectra at very large scales 
$k_{1} < 1/\delta_{\infty}$ are kinematically constrained to be independent of
 $k_1$ and therefore of the form  $E_{11}(k_{1}) \approx C_{\infty} u_{\tau}^{2} \delta$.
This new model predicts that the new range  $1/\delta_{\infty} < k_{1} <1/\delta_{*}$
where $E_{11}(k_{1}) \approx (k_{1}\delta)^{-m}$ is present only for $y$ smaller 
than a wall distance $y_*$ above which the original Townsend-Perry model is valid 
without alterations.\

 By integration of $E_{11}(k_{1})$ we obtain the modified model for the  mean square 
streamwise fluctuating velocity in the range $\nu/u_{\tau} \ll y \le y_{*}=\delta A^{1/p} Re_{\tau}^{-q/p}$:
\begin{equation}
{1\over 2} \overline{u'^{2}}(y)/u_{\tau}^{2} \approx C_{s0} - C_{s1}\ln (\delta/y) -C_{s2} (y/\delta)^{p(1-m)}Re_{\tau}^{q(1-m)}
\label{eq:model}
\end{equation}
with 
\begin{eqnarray}
\label{eq:model1} C_{s0} &=& {C_{0}\over 1-m} + C_{0} \ln B + C_{0}\alpha {q\over p} \ln Re_{\tau} \\
\label{eq:model2} C_{s1} &=& C_{0} (\alpha-1) \\
\label{eq:model3} C_{s2} &=&{m C_{0} A^{m-1}\over 1-m}.
\end{eqnarray}
and where $\alpha$, $B$, $p$ and $q$ are the parameters used in the power 
law formulation of the two scales $\delta_{*}$ and $\delta_{\infty}$ as a 
function of wall distance and Reynolds number:
\begin{equation}
\delta_{*}/\delta = B \; (y/\delta)^{\alpha} Re_{\tau}^{\beta} 
\end{equation}
and 
\begin{equation} 
\delta_{\infty}/\delta_{*} = A \; (y/\delta)^{-p} Re_{\tau}^{-q}
\end{equation}
where $\beta = \alpha \, q/p$ and $B = A^{-\alpha/p}$. This model predicts that the
location $y_{peak}$ of the outer peak or plateau scales as $\delta \, Re_{\tau}^{-q/p}$,
i.e. $y_{peak}^{+} \sim Re_{\tau}^{1-q/p}$. In the region were $y_{*} < y$ equation 
(\ref{eq:logdecay}) holds and completes the model.\\

The parameters of this new model (\ref{eq:model})-(\ref{eq:model3}) have been
evaluated against the Princeton superpipe energy spectra for a large Reynolds
number range ($1\,985<Re_\tau<98\,160$) by \cite{vassilicos15}. It was shown that
the new model is able to reproduce the correct scaling of the outer peak with 
a single set of parameters for all Reynolds numbers. It is important to determine
whether this new model can also account for a variety of high Reynolds number
turbulence boundary layer data and whether it can do it without much variation 
in its defining parameters.

\section{Fits of high Reynolds number TBL data}

In order to test the universality of the model 
(\ref{eq:logdecay}, \ref{eq:model}-\ref{eq:model3}), statistics from several large
Reynolds number turbulent boundary layer experiments have been collected.
As already shown by \cite{hultmark12},  the mean square streamwise fluctuating 
velocity exhibits a clear outer peak for extremely large Reynolds number only.
As very few experimental results are available for $Re_\tau > 40\,000$ and as 
the model is designed and able to capture the outer peak or plateau region even
in the absence of a clear peak, the model is also fitted here on data from 
experiments with smaller Reynolds numbers where the statistics of
$\overline{u'^{2}}$ do not display an outer peak but just a tendency
towards a plateau. The minimum Reynolds number required to fit the
parameters of the model should be such that there is at least a short
range of wall distances where the log decay of $\overline{u'^{2}}$ is
visible.\
In the present contribution, data sets from four turbulent boundary layer
experiments are investigated and compared, covering a range of Reynolds 
numbers from $Re_\tau \simeq 3\,200$ to  $Re_\tau \simeq 72\,000$. The model’s parameters are 
also compared with the original values from the fits of the Princeton superpipe
data by \cite{vassilicos15}.\

The first set of data has been recorded in the turbulent boundary layer wind 
tunnel of Lille which has a test section length of $20.6\, m$ in the stream-wise 
direction (x-direction) and a cross-section 2 $m$ wide and 1 $m$ high. 
Four free-stream velocities were investigated from $U=3\, m/s$ to $U=10\, m/s$
leading to a Reynolds number range from $Re_\tau=3\,200$ to $Re_\tau=7\,000$.
The statistics are detailed in \cite{carlier05}. The hot wire data were recorded
at 19.6 m from the entrance of the test section and the friction velocities
were measured using macro-PIV by \cite{foucaut06}. The second set of data is 
from hot-wire measurements in the large boundary layer wind tunnel at 
University of Melbourne. It has a $27\, m$ long test section of $2 \times 2\, m^2$.
 The statistics of the two lower Reynolds numbers are presented in \cite{marusic15} 
whereas the data for $Re_\tau=19\,000$ were extracted from \cite{marusic10a}.
The third set of data is from the turbulent boundary layer flow physics facility 
(FPF) at the University of New Hampshire (UNH) which is able to reach similar 
Reynolds numbers with a much longer test section ($72\, m$) and a lower free
stream velocity ($U=13.75 \, m/s$). The facility and the hot wire data are
described in \cite{vincenti13}. The facility is an open circuit suction tunnel
that draws from, and discharges to, the atmosphere. The streamwise free-stream 
turbulence  intensity for $U>7 \, m/s$ is less than about 0.3\%.
The fourth and last set of data investigated here are the turbulent boundary
layer and the turbulent pipe data from the Princeton superpipe 
(see \cite{vallikivi15}). A comparison of spectra from both superpipe
experiments over the same range of Reynolds numbers ($3\,000 < Re_\tau < 70\,000$)
 was conducted by \cite{vallikivi14a} leading to 5 scaling regions for the
 turbulent pipe and the turbulent boundary layer.
The data acquired with the Nano Scale Thermal Anemometry Probe (NSTAP)
 are known to be affected by insufficient probe resolution (see Table {\ref{tab:exp}}).
 As in most studies of the authors of the NSTAP data, the present analysis is conducted 
with the data corrected for spatial filtering (see \cite{smits11b}). The correction 
affects mainly the statistics of the two highest Reynolds numbers ($Re_\tau > 40\,000$) for $y^+<300$.
 However the present analysis was also conducted on the raw data to check
 that the quantitative conclusions of the present paper do not change significantly.\\

The data of each experiment are affected by
 different statistical convergence levels and/or slightly different experimental 
conditions. Some important characteristics of the four TBL experiments are summarized in Table \ref{tab:exp}.
For instance, the LML facility produces a slight favorable pressure
 gradient (-0.5 $Pa/m$ at 10 $m/s$) which is not the case for the New Hampshire and the Melbourne facilities.
 Hot wire acquisition time are of the order of $12\,000$ $U_\infty/\delta$ for the 
 Melbourne and LML facilities and 2 to 4 times smaller for the 
New Hampshire facility depending on the Reynolds number. This can explain
 apparent different levels of convergence of some statistics. There are also different
 tripping conditions for the TBL experiments with 
 respect to the size of the boundary layer thickness (see Table \ref{tab:exp}) and in the way
that scaling parameters such as friction velocity ($u_\tau$) and boundary layer
 thickness  ($\delta$) have been determined
 by different techniques. For instance the boundary layer thickness is determined
 as $\delta_{99}$ except for the Melbourne data where $\delta$ is determined from
a modified Coles law of the wall/wake fit to the mean velocity profile. This method 
can overestimate $\delta$ by up to 25\% with respect to $\delta_{99}$ depending on 
the Reynolds number.  The friction velocity $u_\tau$ is either determined by Preston tube
 or Clauser plot. The accuracy of both methods is known to be of the order of few percent.
 The values of $u_\tau$ for Lille experiments are validated by micro-PIV.\\

\begin{figure}
\begin{center}
\begin{minipage}{0.49\columnwidth}
\includegraphics[width=1.0\columnwidth]{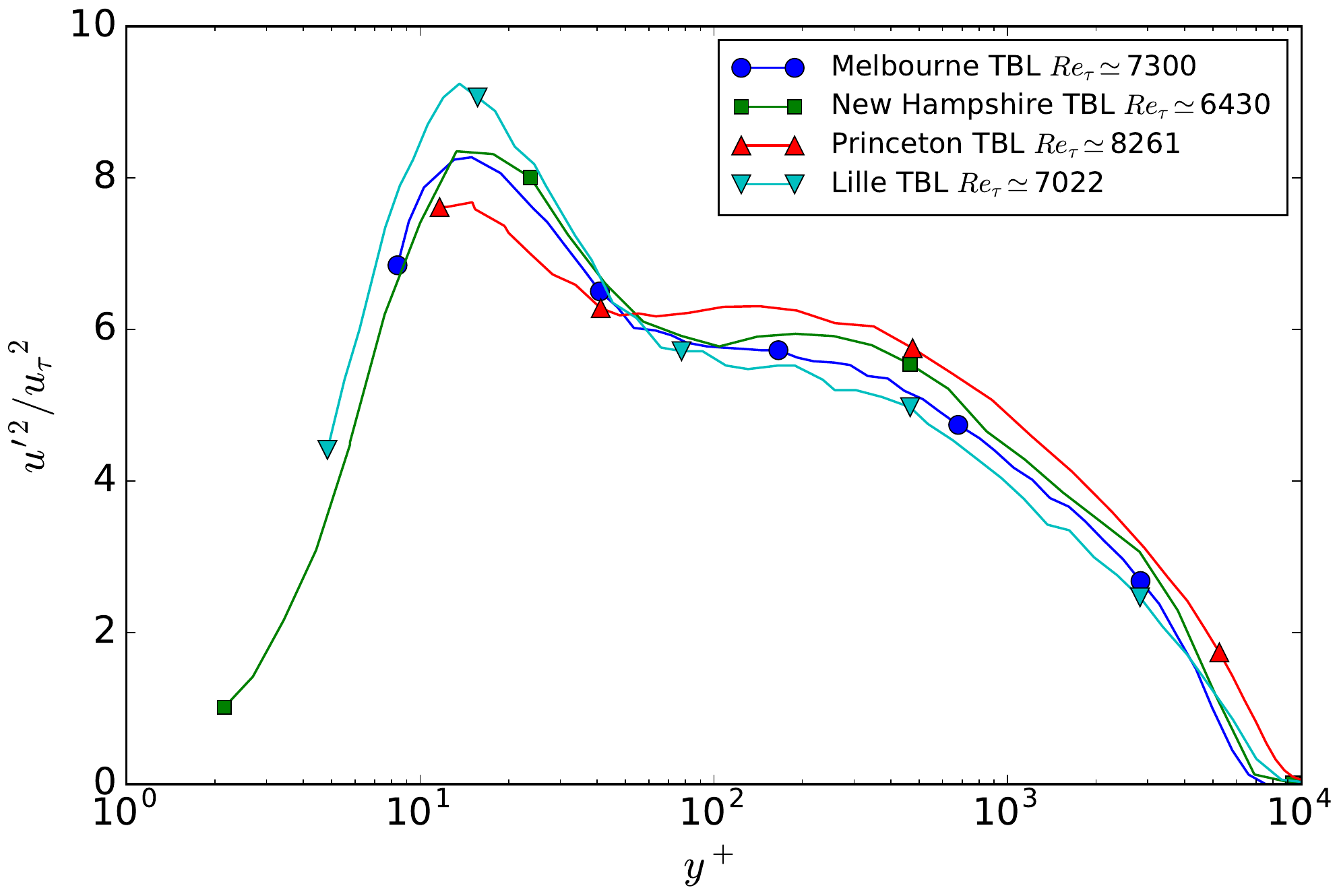}\\
\end{minipage}
\begin{minipage}{0.49\columnwidth}
\includegraphics[width=1.0\columnwidth]{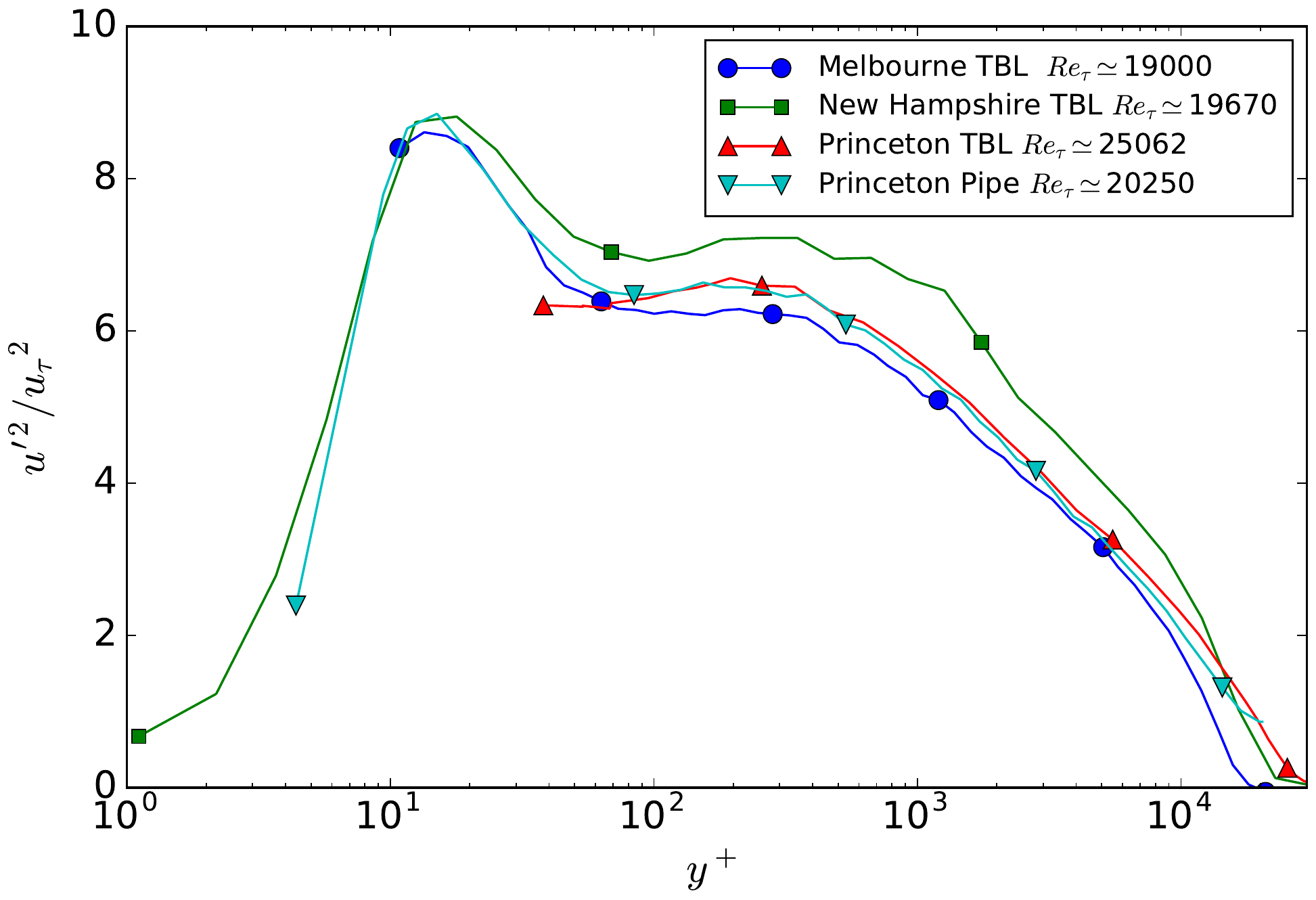}\\
\end{minipage}
\caption{\label{fig:same_re} Comparison of $\overline{u'^{2}}/u_\tau^2$ versus $y^+$ for
 several turbulent boundary layer and pipe flow experiments at two Reynolds numbers: 
a) $Re_\tau \simeq 7\,000$ b)  $Re_\tau \simeq 20\,000$}
\end{center}
\end{figure}

In figure \ref{fig:same_re} we plot, for comparison, mean square streamwise 
fluctuating velocity profiles from different experiments at two similar 
Reynolds numbers. As both the boundary layer thickness and the length of 
the hot wires are different, the comparison of the results in the region 
of the first peak ($y^+ < 30$) is not significant, especially at the highest 
Reynolds number. Due to a different definition of the boundary layer thickness,
 the Reynolds number for the Melbourne data may be over-estimated which may explain the slightly lower
values of mean square streamwise fluctuating velocity profiles.
 However, taking into account the slightly different Reynolds
number values and experimental conditions in each plot of figure \ref{fig:same_re} and the uncertainty 
in the friction velocity evaluations, it can be said that the four experiments 
exhibit comparable overall behaviours. The values of  $\overline{u'^{2}}$  in 
the New Hampshire TBL are however slightly higher than the average of the 
three other experiments for the higher of the two Reynolds numbers 
(figure \ref{fig:same_re}b). The log decay (\ref{eq:logdecay}) is clearly 
defined only at the highest Reynolds number ($Re_\tau \simeq 20\,000$) except
for the New Hampshire TBL data for which the outer peak is more pronounced 
and extends to higher $y^+$.\\

\begin{table}
\caption{\label{tab:exp} Parameters of the four turbulent boundary layer experiments. The boundary layer thickness $\delta$ corresponds to $\delta_{99}$ except for Melbourne TBL where $\delta$ is determined from a modified Coles law of the wall/wake fit to the mean velocity profile. $\ell^+$ is the sensor length in wall unit. }
\begin{center}
\begin{tabular}{|l|c|c|c|l|}
\multicolumn{5}{c}{}\\
\hline
Experiments      & $Re_\tau$  & $\delta$  & $\ell^+$ & Tripping conditions \\
\hline
\multirow{3}{*}{\begin{tabular}{l} Lille TBL \\ \cite{carlier05}  \end{tabular}} & 3196 & 0.319  & 6.1  & \multirow{3}{*}{\begin{tabular}{l} Grid  5 mm thickness and 10 cm spacing \\ on 2 m at the entrance of the test section \end{tabular}} \\
                           & 5006 & 0.298  & 8.4  &  \\
                           & 7022 & 0.304  & 11.5 &  \\
\hline
\multirow{3}{*}{\begin{tabular}{l} Melbourne TBL \\ \cite{marusic10a} \\ \cite{marusic15} \end{tabular}} & 10000 &  0.226 & 23.9 &  \multirow{3}{*}{\begin{tabular}{l}Grit P40 Sandpaper 154 mm long located\\  at the entrance of test section ($Re_\tau\!\!=\!\!10000$)  \end{tabular}} \\
                               & 13600 &  0.315 & 22   &  \\
                               & 19000 &  0.303 & 22   &  \\
\hline
\multirow{3}{*}{\begin{tabular}{l} New Hampshire TBL \\ \cite{vincenti13} \end{tabular}} & 10770 & 0.736 & 14.6 &  \multirow{3}{*}{\begin{tabular}{l} 6mm threaded rod 1 mm above the surface \\ at 1.4 m from the test section entrance \end{tabular}} \\
                                   & 15480 & 0.717 & 22   & \\
                                   & 19670 & 0.688 & 28.6 & \\
\hline
\multirow{4}{*}{\begin{tabular}{l} Princeton TBL \\ \cite{vallikivi14a} \end{tabular}}     & 8261  & 0.0283 & 17   & \multirow{4}{*}{\begin{tabular}{l} 1mm square wire 76mm from leading edge  \\ 1.82 m upstream the measurement location.\end{tabular} } \\
                                   & 25062 & 0.0257 & 29   & \\
                                   & 40053 & 0.0257 & 47   & \\
& 72526 & 0.0291 & 75   & \\
\hline
\end{tabular}
\end{center}
\end{table}

As a first test, the model (\ref{eq:logdecay}), (\ref{eq:model})-(\ref{eq:model3}) 
was fitted to each Reynolds number data of each dataset independently.
This results in a set of optimal parameters defined in table \ref{tab:param}.
The two regions $y<y_{*}$ and $y>y_{*}$  are fitted at the same time but as the 
parameters $A_1$ and $B_1$ involved in the log decay region $y>y_*$ are fitted 
independently of $C_0$, the two fitted parts do not perfectly match.
The extent of the region with the logarithmic decay (\ref{eq:logdecay}) is not
a priori certain. In the present case a conservative but widely accepted upper
bound $y<0.15\delta$ was used. This choice leads to a rather short logarithmic
region for the lower Reynolds numbers investigated here and consequently a
substantial uncertainty in the determination of the two constants $B_1$ and $A_1$. 
The two parts of the model  as well as the wall distance $y_*$ where the two 
models merge are fitted simultaneously using a L-BFGS-B algorithm developed by
\cite{byrd95} for the solution of the optimisation problem.

The first conclusion is that the model is able to fit the complete dataset 
with parameter values which do not vary significantly with Reynolds number and
from one experiment to the other. The two exponents $p$ and $q$ control the 
wall distance of the outer peak or plateau as well as the related Reynolds 
number and wall-distance dependencies of the new spectral range extent
$\delta_{\infty}/\delta_{*}$. The values obtained for $p$ and $q$ appear relatively constant across Reynolds numbers 
and across TBL experiments. They appear also fairly close to those obtained by 
\cite{vassilicos15} for the Princeton high Reynolds number turbulent pipe flow.
These values of $q$ and $p$ suggest an outer peak position scaling like
$y_{peak}^{+} \sim Re_{\tau}^{2/3}$ (following section 2, $1-q/p \simeq 0.61$,
 which is close to $2/3$) and the fact that $\alpha$
 is always larger than 1 suggests that an outer peak does indeed exist 
(see \cite{vassilicos15} for an explanation of this point). However the ratio 
$q/p$ seems to decrease slightly for the  Princeton TBL with increasing 
Reynolds number. The parameter $m$ which, in the model, controls the slope of 
the $k_{1}^{-m}$ spectrum in the new range $1/\delta_{\infty} < k_{1} < 1\delta_{*}$
is extremely stable as compared, for instance, to $\alpha$.
 However, the two values appear to be correlated so that a small
variation of the former can be compensated by the later. The parameter $C_0$ 
which controls the amplitude of the outer peak does not vary too much and 
remains close to a mean value of 1.33. The larger values of $C_0$ for the  
New Hampshire data are linked to the more pronounced outer peak as already 
noticed in Fig. \ref{fig:same_re}. The parameters $A_1$ and $B_1$ of the 
log decay model (\ref{eq:logdecay}) are only significant at the highest Reynolds
numbers investigated where this log decay region is clearly defined.
They are also dependent on the lower bound on $y$ chosen to fit the model 
(\ref{eq:logdecay}).
 In our model, the log decay region starts at the wall distance $y_*$ which 
was used to fit the two parameters $A_1$ and $B_1$. The average values of these 
two parameters keeping only the 7 cases with $Re_\tau > 12\,000$ are 
$A_1 \simeq 1.35$ and $B_1 \simeq 1.45$. The average slope $A1$ is close to the value reported by 
\cite{marusic13}  with a different definition of the fitting lower bound and
 taking into account their 95\% confidence estimated error bars.\\

\begin{table}
\caption{\label{tab:param} Best values of the model parameters (\ref{eq:logdecay}, \ref{eq:model}-\ref{eq:model3}) for the fit of the four turbulent boundary layer datasets. }
\begin{center}
\begin{tabular}{lccccccccc}
\\
\hline
Experiments                           & $Re_\tau$  &   m   &  p    &   q   &   A     & $\alpha$ & $C_0$ &  $B_1$  &  $A_1$    \\ \hline
\multirow{3}{*}{\begin{tabular}{l} Lille TBL \\ \cite{carlier05}  \end{tabular}}     &    3193   & 0.38  & 2.13 &  0.77  &   1.12  &   1.26   &  1.23 &  1.70  &  1.15      \\ 
                                                                                     &    5006   & 0.38  & 2.14 &  0.76  &   1.10  &   1.33   &  1.28 &  1.70  &  1.18      \\ 
                                                                                     &    7022   & 0.38  & 2.16 &  0.77  &   1.50  &   1.03   &  1.27 &  1.69  &  1.18      \\ \hline
\multirow{4}{*}{\begin{tabular}{l} Melbourne TBL \\ \cite{marusic10a} \\ \cite{marusic15} \end{tabular}} &    7172   & 0.38  & 2.20 &  0.77  &   1.57  &   1.09   &  1.35 &  1.70  &  1.29      \\ 
                                                                                     &    10000  & 0.38  & 2.13 &  0.76  &   1.15  &   1.20   &  1.36 &  1.70  &  1.30      \\ 
                                                                                     &    13600  & 0.35  & 2.30 &  0.79  &   0.92  &   1.13   &  1.32 &  1.50  &  1.30      \\ 
                                                                                     &    19000  & 0.38  & 2.11 &  0.79  &   1.04  &   1.22   &  1.28 &  1.59  &  1.23      \\ \hline
\multirow{3}{*}{\begin{tabular}{l} New Hampshire TBL \\ \cite{vincenti13} \end{tabular}} &10770  & 0.35  & 2.30 &  0.75  &   1.80  &   1.10   &  1.56 &  1.40  &  1.65      \\ 
                                                                                     &    15740  & 0.38  & 2.12 &  0.78  &   1.06  &   1.32   &  1.47 &  1.70  &  1.59      \\ 
                                                                                     &    19670  & 0.38  & 2.12 &  0.78  &   1.03  &   1.33   &  1.51 &  1.70  &  1.64      \\  \hline

\multirow{4}{*}{\begin{tabular}{l} Princeton TBL \\ \cite{vallikivi14a} \end{tabular}} &   8261  & 0.38  &  2.10 &   0.81 &  0.99 &  1.28  &  1.37 &  1.70 &  1.43 \\ 
                                                                                     &    25062  & 0.38  &  2.09 &   0.81 &  0.99 &  1.18  &  1.30 &  1.43 &  1.29 \\ 
                                                                                     &    40053  & 0.38  &  2.05 &   0.84 &  0.95 &  1.13  &  1.21 &  1.24 &  1.24 \\ 
&    72526  & 0.39  &  2.00 &   0.89 &  0.90 &  1.20  &  1.17 &  1.13 &  1.16 \\
\hline

\end{tabular}
\end{center}
\end{table}

\begin{figure}
\begin{center}
\begin{minipage}{0.49\columnwidth}
\includegraphics[width=1.03\columnwidth]{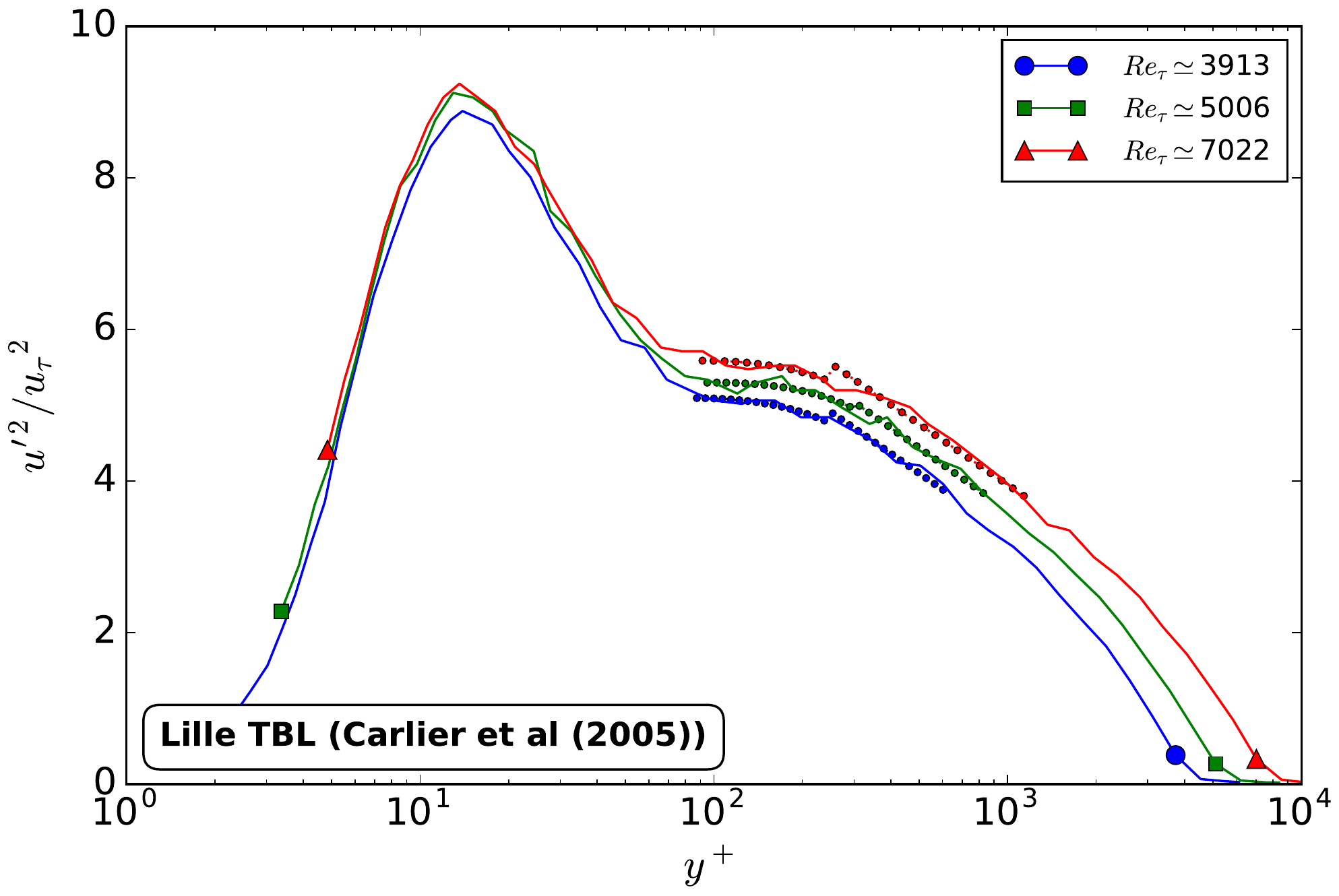}\\
\includegraphics[width=1.0\columnwidth]{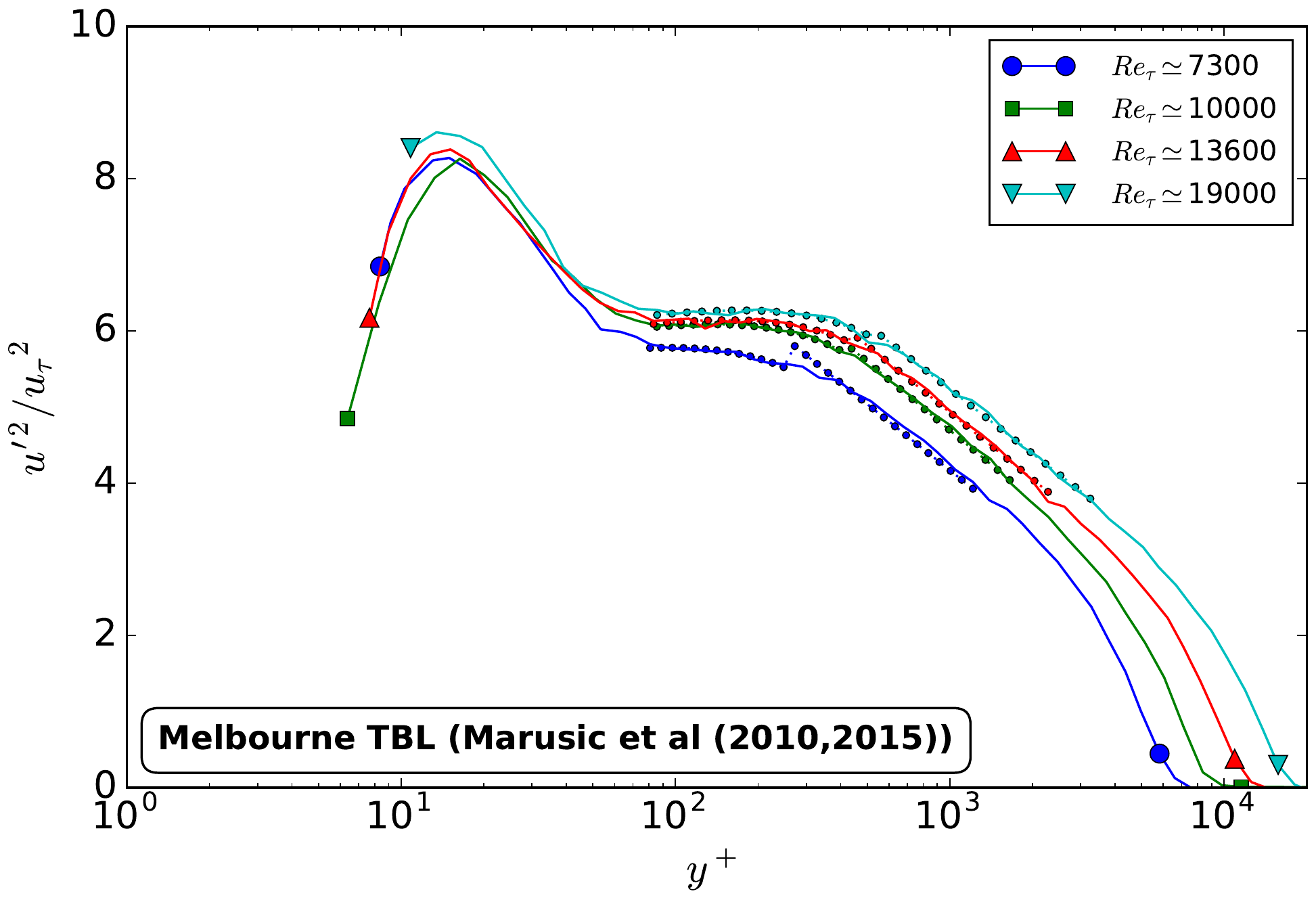}\\
\end{minipage}
\hfill
\begin{minipage}{0.49\columnwidth}
\includegraphics[width=1.0\columnwidth]{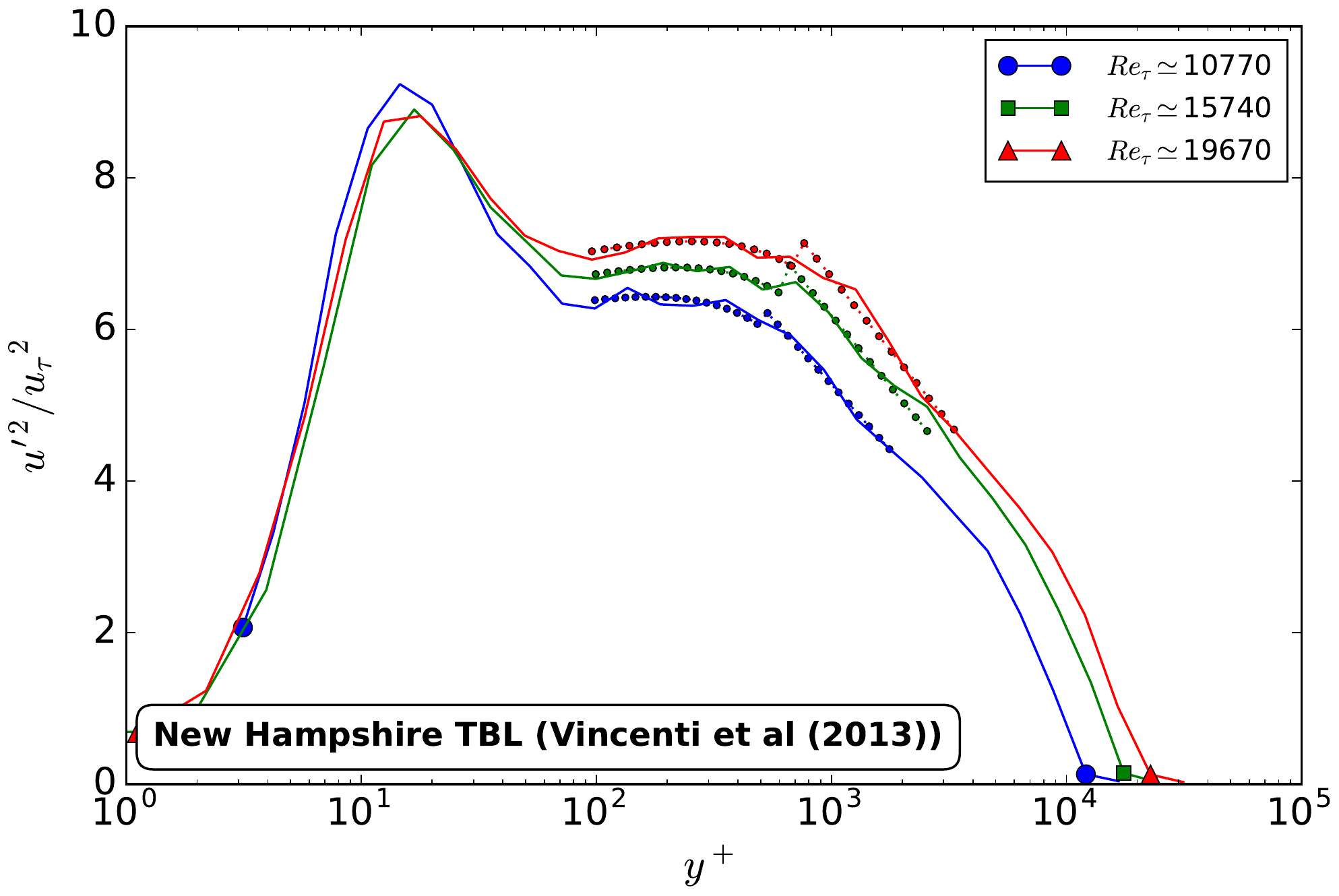}\\
\includegraphics[width=1.0\columnwidth]{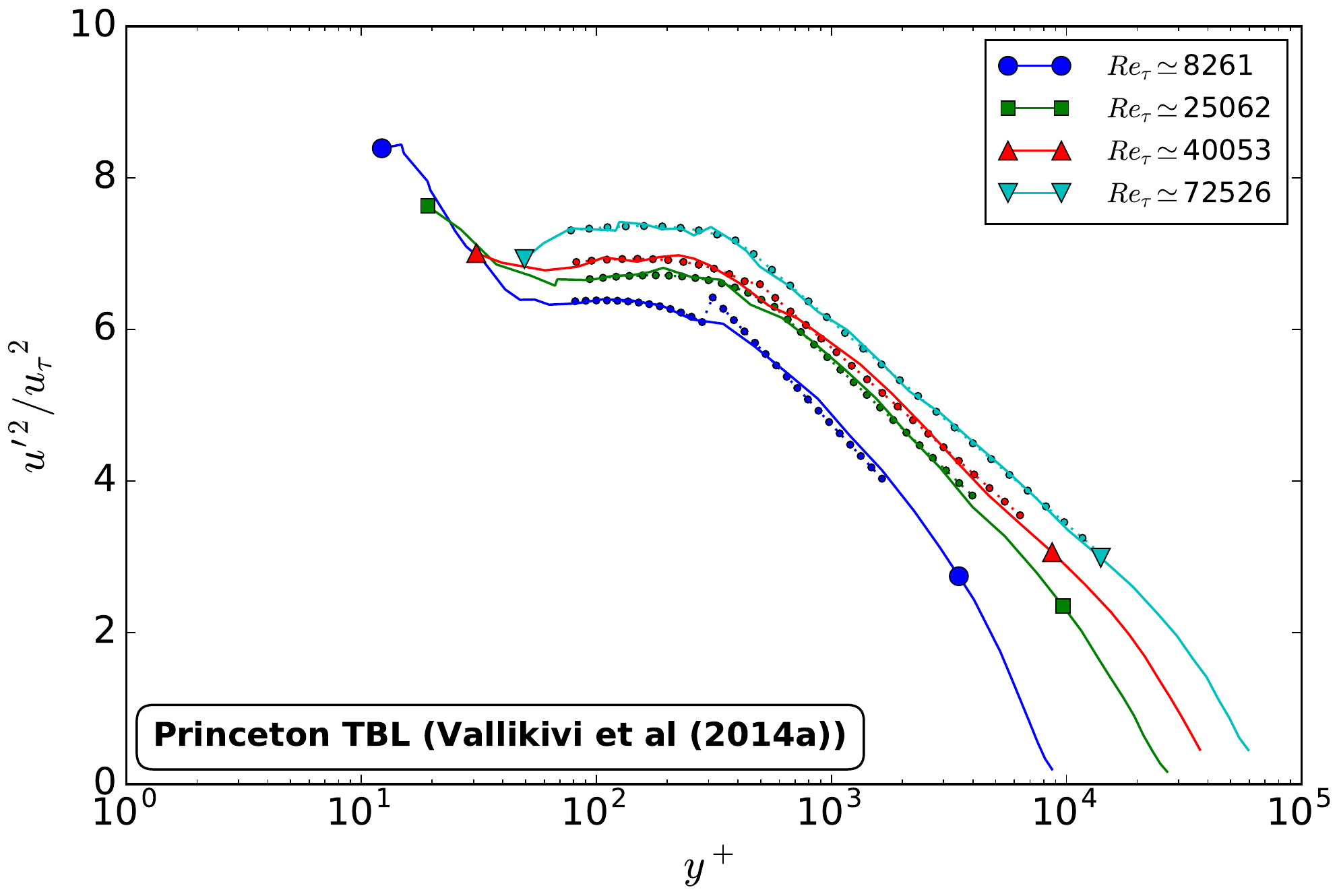}\\
\end{minipage}
\begin{minipage}{0.49\columnwidth}
\includegraphics[width=1.0\columnwidth]{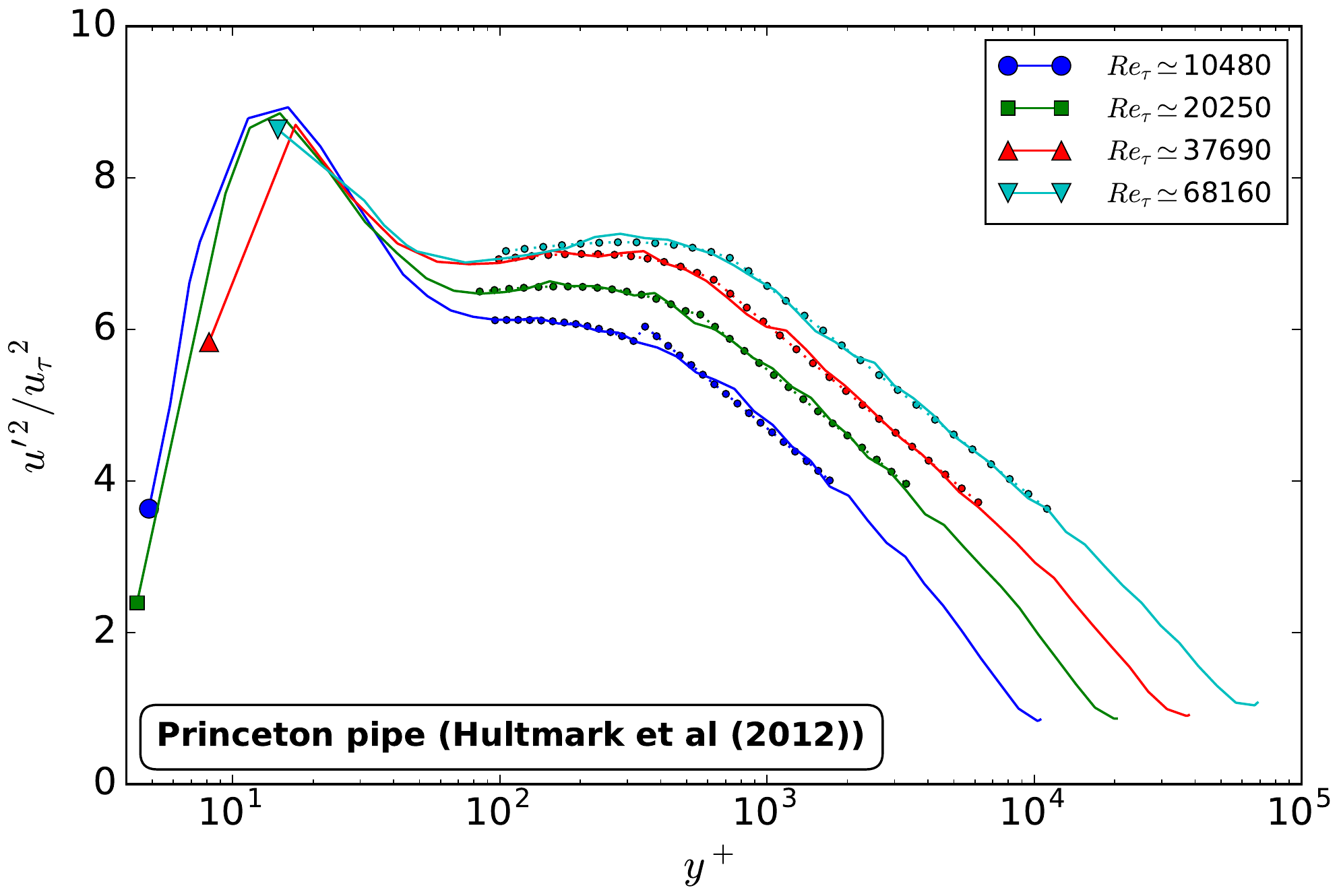}\\
\end{minipage}
\begin{minipage}{0.49\columnwidth}
\hfill
\end{minipage}
\caption{\label{fig:fit} Fit of each experiment dataset with the model (\ref{eq:logdecay}, \ref{eq:model}-\ref{eq:model3}) using 4 variable parameters ($q$, $C_0$, $B_1$, $A_1$) and the remaining ones fixed at $m=0.38$, $p=2.1$, $A=1.0$ and $\alpha=1.2$. }
\end{center}
\end{figure}

As a second test of robustness of the model, the data of the four TBL
experiments were fitted with the  model whilst keeping a constant value for 
four parmeters ($m=0.38$, $p=2.1$, $A=1.0$ and $\alpha=1.2$).
 The values of these four parameters are taken as the approximate
average of the first estimation when all parameters are fitted. We choose to 
adjust the four remaining parameters $q$, $C_0$, $B_1$ and $A_1$ for two reasons.
Firstly, the model is sensitive to the ratio $p/q$ which controls the location 
of the outer peak and more importantly the wall distance $y_*$ associated 
to the merging of the two parts of the model. 
Secondly, the other three parameters $C_0$, $B_1$ and $A_1$ are associated with
the amplitude of the outer peak and the slope and level of the log-decay region.
The values of $A_1$ and $B_1$ in (\ref{eq:logdecay}) vary from one experiment to 
the other in particular because this region is well established only at extremely
large Reynolds numbers but also because there is no clear consensus on the exact 
bounds on $y$ where this log-decay region exists. The four fitted parameters are 
given in table \ref{tab:param2} and the resulting fits for the four TBL data 
(and for the Princeton turbulent pipe data for comparison) are shown in 
Fig. \ref{fig:fit}. The first part of the model (i.e. equations
 (\ref{eq:model})-(\ref{eq:model3}) for $80 \nu/u_{\tau} \lesssim y < y_*$) which results from the addition 
of the new spectral range $1/\delta_{\infty} < k_1 < 1/\delta_*$ is able to return 
good fits for all the experiments and all the Reynolds numbers investigated.
The parameters $C_0$ and $q$ which control the amplitude and location of the
outer peak (or plateau for $7\,000 < Re_\tau < 20\,000$) take values similar
to the previous fit (Table \ref{tab:param}); the rms difference on these two 
parameters is only about 3\% between Tables  \ref{tab:param} and \ref{tab:param2}. 
The variations in the values of the log decay parameters $B_1$ and $A_1$ with 
respect to Table \ref{tab:param} are due to the slight changes of the lower 
bound $y_*$. Finally, the parameters $q$, $C_0$, $A_1$ and $B_1$ are also given 
for the Princeton turbulent pipe data at four Reynolds numbers. At similar 
Reynolds numbers, the differences in the values of these four parameters between 
the two types of flow (TBL and turbulent pipe flow) are comparable to the 
differences in their values from one TBL experiment to the other.\\

\begin{table}
\caption{\label{tab:param2} Best values of the model parameters (\ref{eq:logdecay},
\ref{eq:model}-\ref{eq:model3}) for the fit of the four turbulent boundary layer 
datasets and the Princeton superpipe data. The fit was performed fixing four of 
the parameters ($m=0.38$, $p=2.1$, $A=1.0$ and $\alpha=1.2$).}
\begin{center}
\begin{tabular}{lccccccccc}
\\
\hline
Experiments                           & $Re_\tau$  &  $q$   &  $C_0$ &  $B_1$  &  $A_1$    \\ \hline
\multirow{3}{*}{\begin{tabular}{l} Lille TBL \\ \cite{carlier05}  \end{tabular}}                         &  3193   & 0.71 & 1.26 & 1.70  & 1.17  \\ 
                                                                                                         &  5006   & 0.70 & 1.29 & 1.70  & 1.19  \\ 
                                                                                                         &  7022   & 0.78 & 1.23 & 1.70  & 1.15  \\ \hline
\multirow{4}{*}{\begin{tabular}{l} Melbourne TBL \\ \cite{marusic10a} \\ \cite{marusic15} \end{tabular}} &  7172   & 0.78 & 1.28 & 1.70  & 1.25  \\ 
                                                                                                         &  10000  & 0.74 & 1.36 & 1.70  & 1.30  \\
                                                                                                         &  13600  & 0.76 & 1.31 & 1.58  & 1.27  \\
                                                                                                         &  19000  & 0.78 & 1.28 & 1.59  & 1.23  \\ \hline
\multirow{3}{*}{\begin{tabular}{l} New Hampshire TBL \\ \cite{vincenti13} \end{tabular}}                 &  10770  & 0.70 & 1.48 & 1.70  & 1.50  \\ 
                                                                                                         &  15740  & 0.70 & 1.53 & 1.70  & 1.63  \\ 
                                                                                                         &  19670  & 0.70 & 1.58 & 1.70  & 1.67  \\  \hline
\multirow{4}{*}{\begin{tabular}{l} Princeton TBL \\ \cite{vallikivi14a} \end{tabular}}                   &   8261  & 0.76 & 1.40 & 1.70  & 1.44  \\ 
                                                                                                         &  25062  & 0.81 & 1.30 & 1.44  & 1.28  \\ 
                                                                                                         &  40053  & 0.90 & 1.20 & 1.34  & 1.20  \\ 
                                                                                                         &  72526  & 0.95 & 1.17 & 1.13  & 1.16  \\  \hline
\multirow{4}{*}{\begin{tabular}{l} Princeton Pipe \\ \cite{hultmark12} \end{tabular}}                    &  10480  &  0.78 & 1.31 & 1.70  & 1.28  \\ 
                                                                                                         &  20250  &  0.79 & 1.33 & 1.69  & 1.26  \\ 
                                                                                                         &  37690  &  0.82 & 1.30 & 1.34  & 1.29  \\ 
&  68160  &  0.83 & 1.27 & 1.43  & 1.22  \\
\hline

\end{tabular}
\end{center}
\end{table}

The evidence obtained from the present analysis points to a wide applicability 
of the model (\ref{eq:model})-(\ref{eq:model3}) and
(\ref{eq:logdecay}), with parameter values which do not vary much from one flow to 
the other (a wide range of Reynolds numbers $Re_\tau$ from  $7\,000$ to $70\,000$,
two types of wall-bounded turbulent flows and four different installations). This
motivates us to propose a best set of values for all eight parameters 
of the model which can be used for predictive purposes, for example when 
extrapolating to near-wall turbulence at even higher Reynolds numbers. However 
very few data are available to test such extrapolation. An example of such rare 
data was obtained from highly documented measurements performed at SLTEST, an 
atmospheric facility in the Great Salt Lake Desert. Several sets of experiments
were conducted in this facility but we focus only on the most reliable ones, 
the sonic anemometer data from \cite{hutchins12} and the hot wire and sonic 
anemometer data from \cite{metzger07}. The first set covers only part of the 
log decay region while both hot wire and sonic anemometer data are available 
from the second set covering the two specific regions of the present model. 
As parameters vary a little with Reynolds number when the Reynolds number is 
not high enough, the parameters used for the extrapolation were computed as 
their  average for TBL with  $Re_\tau \ge 12\,000$ (Table \ref{tab:param}) only. 
This leads to $A_1 \simeq 1.35$, $B_1 \simeq 1.45$,  $A \simeq 1.0$,  
$\alpha \simeq 1.21$, $p \simeq 2.1$, $q \simeq 0.81$, $m \simeq 0.38$ and 
$C_0 \simeq 1.32$. However, as the continuity of the model at $y=y_*$ imposes  
$A_1=B_1=C_0$  (see \cite{vassilicos15} for full explanations) we choose a 
compromise value of $A_1=B_1=C_0=1.4$ keeping the other parameters to their 
average value.

\begin{figure}
\begin{center}
\includegraphics[width=0.85\columnwidth]{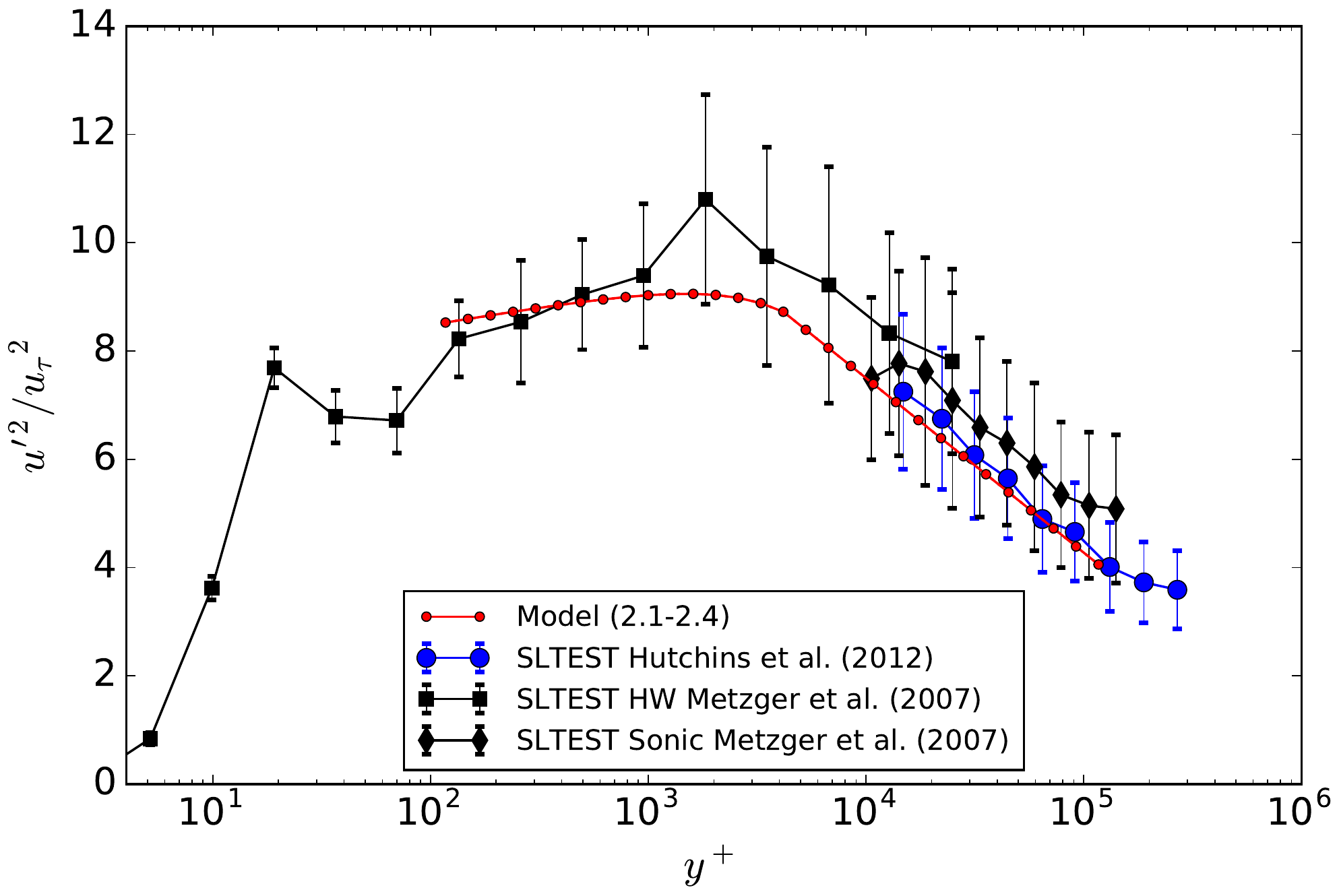}\\
\end{center}
\caption{\label{fig:2} Comparison  of $\overline{u'^{2}}/u_\tau^2$ for  the measurements campaign of the SLTEST experiments with the model (\ref{eq:logdecay})-(\ref{eq:model}-\ref{eq:model2}) using the averaged parameters obtained from previous fit for the highest Reynolds numbers ($m=0.38$, $p=2.1$, $q=0.81$, $A=1.0$, $\alpha=1.21$, $C_0=1.4$, $B_1=1.4$, $A_1=1.4$). The Reynolds numbers based on boundary layer thickness were estimated to 60 millions and 80 millions of the \cite{hutchins12} and \cite{metzger07} corresponding to $Re_\tau \simeq 770\,000$ and $Re_\tau \simeq 780\,000$ respectively.}
\end{figure}

These average values for $p$, $m$, $\alpha$ and $A$ are also the values chosen 
for our second fit (Table \ref{tab:param2}). The two sets of SLTEST
data as well as our model parameters and prediction are shown in Fig. \ref{fig:2}. The values 
chosen for $A_1$ and $B_1$ lead to a prediction of the log decay region in 
general agreement with the two sets of data taking into account the error bars 
proposed by the authors. The model predicts an “outer peak” in global agreement
with the hot wire data of  \cite{metzger07} given the accuracy of the experimental
results. In fact, the outer peak returned by the model looks more like a plateau 
with a weak slope. Putting aside the point of Metzger et al. at $y^+ = 2000$ 
which is potentially questionable, the global shape and level predicted by the
extrapolation of our model to these extremely high Reynolds numbers agrees 
reasonably well with the SLTEST data down to $y^+ = 150$. It would be of interest
to have more reliable measurements of the inner peak in such a flow to see how 
the inner and outer peaks connect.\\

\section{Conclusions}

Several studies of high Reynolds number turbulent pipe, channel and zero-pressure gradient boundary layer flows have shown that a region of high turbulence develops further away from the wall than the well known near-wall turbulence peak (see \cite{hultmark12}, \cite{morrisson04},\cite{marusic10a}). In a recent paper, \cite{vassilicos15} developed a model which, based on a new reading of turbulence spectra, combines this outer high turbulence region with the Townsend-Perry log-decay of the rms turbulence velocity profiles. The model is able to account for this new turbulence region and its Reynolds number dependence from a newly identified low wavenumber range of the energy spectrum with attached eddy physical significance. The model was validated by \cite{vassilicos15} on the exceptionally high Reynolds number data of the Princeton superpipe.

It is of course important to establish how widely this model holds and to see whether its validity can be expanded to turbulent boundary layers. For that purpose, we have collected data from the few facilities around the world which can provide turbulent boundary layer measurements at Reynolds numbers sufficiently high to be relevant. Thanks to the kindness of colleagues from Melbourne, New Hampsire and Princeton, a fairly comprehensive dataset could be assembled covering a range of $Re_\tau$ from approximately $3,000$ to $70,000$. After checking the coherence of the dataset, two different fits where tried: one where all the parameters were free to be fitted by the optimisation procedure and one where four of the parameters where chosen a priori on the basis of the first fit and four were fitted. Both approaches give a reasonable prediction of the entire dataset and values of the parameters which are in good agreement. Furthermore, these values are fairly close to the ones obtained by \cite{vassilicos15} for the Princeton superpipe data.

These encouraging results allow us to propose a set of parameters to be used to predict high Reynolds number near wall turbulence. We therefore tried the predictive power of our model on the very highest Reynolds number data available: those of the SLTEST facility in the Salt Lake desert in USA. Two datasets from two different teams (Hutchins et al 2012 and Metzger et al 2007) where selected as the most representative of this Salt Lake desert flow. The results show that the prediction of our model is in good agreement with the measurements within measurement accuracy. Our model can therefore be used to extrapolate the near wall turbulence intensity distribution to very high Reynolds numbers and may even provide a means to improve near-wall LES predictions until a better approach becomes available. It is interesting to note that at Reynolds numbers as high as those of the SLTEST facility ($Re_\tau = 780,000$) the model does not yield a very marked peak but rather a plateau with a weak slope.

\section{Acknowledgements}
The authors gratefully acknowledge Ivan Marusic, Joe Klewicki and Alexander Smits for kindly providing us with their experimental data.  This work was supported by ``Campus International pour la S\'ecurit\'e et l'Intermodalit\'e des Transports, la R\'egion Nord-Pas-de-Calais, l'Union Europ\'eenne, la Direction de la Recherche, Enseignement Sup\'erieur, Sant\'e et Technologies de l'Information et de la Communication et le Centre National de la Recherche Scientifique''. J.C.V. acknowledges the support of an ERC advanced grant (2013-2018).


\end{document}